\documentclass[conference]{IEEEtran}
\IEEEoverridecommandlockouts
\usepackage{cite}
\usepackage{amsmath,amssymb,amsfonts}
\usepackage{algorithmic}
\usepackage{graphicx}
\usepackage{textcomp}
\usepackage{xcolor}
\usepackage{url}
\usepackage{booktabs}
\usepackage{amsfonts}
\usepackage{siunitx}
\usepackage{multirow}
\usepackage{enumitem}
\usepackage{subcaption}
\usepackage[breaklinks=true,bookmarks=false]{hyperref}
\def\BibTeX{{\rm B\kern-.05em{\sc i\kern-.025em b}\kern-.08em
    T\kern-.1667em\lower.7ex\hbox{E}\kern-.125emX}}
\def\eg{\textit{e.g.}}   
\def\ie{\textit{i.e.}}   
\def\x{{\mathbf x}}
\begin{document}

\makeatletter
    \newcommand{\linebreakand}{%
      \end{@IEEEauthorhalign}
      \hfill\mbox{}\par
      \mbox{}\hfill\begin{@IEEEauthorhalign}
    }
    \makeatother
\title{Exposing AI-Synthesized Human Voices Using Neural Vocoder Artifacts
}
\author{Chengzhe Sun$^{1}$, Shan Jia$^{1}$, Shuwei Hou$^{1}$, Ehab AlBadawy$^{2}$, Siwei Lyu$^{1}$\\
$^{1}$ Department of Computer Science and Engineering, University at Buffalo, SUNY, Buffalo, USA \\
$^{2}$ Department of Electrical and Computer Engineering, University at Albany, SUNY, Albany, USA \\}
\maketitle

\begin{abstract}
   The advancements of AI-synthesized human voices have introduced a growing threat of impersonation and disinformation. It is therefore of practical importance to develop detection methods for synthetic human voices. This work proposes a new approach to detect synthetic human voices based on identifying artifacts of {\em neural vocoders} in audio signals. A neural vocoder is a specially designed neural network that synthesizes waveforms from temporal-frequency representations, \eg, mel spectrograms. The neural vocoder is a core component in most DeepFake audio synthesis models. Hence the identification of neural vocoder processing implies that an audio sample may have been synthesized. To take advantage of the vocoder artifacts for synthetic human voice detection, we introduce a multi-task learning framework for a binary-class RawNet2 model that shares the front-end feature extractor with a vocoder identification module. 
   We treat the vocoder identification as a pretext task to constrain the front-end feature extractor to focus on vocoder artifacts and provide discriminative features for the final binary classifier. Our experiments show that the improved RawNet2 model based on vocoder identification achieves an overall high classification performance on the binary task\footnote{Dataset and codes will be available at \href{https://github.com/csun22/LibriVoc-Dataset}{\url{https://github.com/csun22/LibriVoc-Dataset}.}}.
\end{abstract}


\section{Introduction}
\label{sec:intro}

Recent years have seen the proliferation of synthetic media, riding the waves of the rapid advancement of AI technologies, in particular, deep learning.  These synthetic media are more commonly known as the ``DeepFakes,'' a portmanteau of deep learning and fake media. State-of-the-art AI media synthesis methods can now create highly realistic still images and videos that challenge the viewer's ability to distinguish them from the real media~\cite{DeepFake:survey:acmcs2021}. While the AI-synthesized still images and videos are currently in the spotlight of public attention, synthetic human voices have also undergone considerable developments and are reaching unprecedented perceptual quality and generation efficiency. These AI-synthesized human voices can facilitate new capacities in voice-based user interfaces for intelligent home assistants and wearable devices and can be used to help patients whose speech abilities are damaged by strokes or  Amyotrophic Lateral Sclerosis (ALS) to gain back voices. However, synthetic human voices could also be misused for deceptions. In one recent incident, a scammer used an AI-synthesized voice to impersonate the CEO of a UK company in a phone call, and misled an employee to wire transfer a substantial amount of money to the scammer’s bank account \cite{forbes}. 

While the detection of AI-synthesized still images and videos have been avidly studied in recent years \cite{lyu_icmew20}, methods to detect synthetic human voices have received relatively less attention and are underdeveloped. This is because audio signals have different characteristics that hinder the direct application of image-based detection methods. Existing detection methods usually examine signal statistical features that are particular to audio signals, for instance, the work in \cite{albadawy2019detecting} compares the higher-order statistics in the bi-spectral domain that capture the local phase inconsistencies in synthetic voices. 

In this work, we propose a new approach to detect synthetic human voices based on the artifacts introduced by the {\em neural vocoders} used in the generation process. A neural vocoder is a special-purpose neural network that synthesizes audio waveforms from temporal-frequency representations such as mel spectrograms. Because neural vocoders are the last step in most AI-based audio synthesis models, it is implausible that real audio signals will be processed with neural vocoders, therefore, they can provide cues to expose synthetic human voices. 


Hence, the foremost objective of our work is to highlight the distinct signal artifacts left by neural vocoders in synthetic audio signals. 
To explore the artifacts of vocoders, we first construct a dataset entitled {\em LibriVoc}, which aims to control other factors and probes for the vocoder signature only and contains evenly distributed data for various vocoders. 
\begin{figure*}[t]
    \centering
    \includegraphics[width=1.0\textwidth]{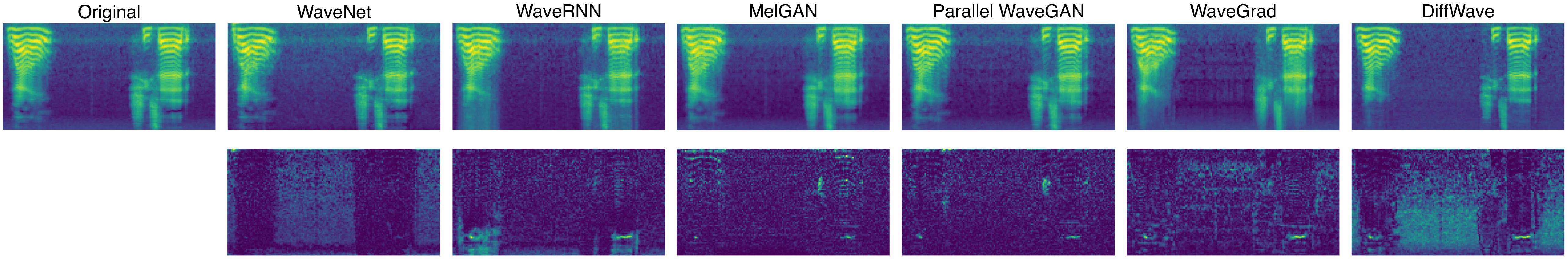}
    ~\vspace{-2.0em}
    \caption{\small The artifacts introduced by the neural vocoders to a voice signal. We show the mel spectrogram of the original (top left) and the self-vocoded voice signal (top right five). Their differences corresponding to the artifacts introduced by the vocoder are shown at the bottom. }
    ~\vspace{-2.5em}
    \label{fig:sig_diff}
\end{figure*}
We use six neural vocoders in creating the LibriVoc dataset to reflect the diversity in the architecture and mechanisms of neural vocoders. Because the ``self-vocoding'' samples are sourced from the same original audio signals, they highlight the artifacts introduced by the vocoders. Fig. \ref{fig:sig_diff} shows the differences in the mel spectrogram of one original voice and its self-vocoded voice signals. Visible artifacts introduced by different neural vocoder models can be observed, which serve as the basis of our detection algorithm. While these artifacts are subtle to visualize, this work demonstrates that they can be captured by a trained classifier. 

To take advantage of the vocoder artifacts in the detection of synthetic human voices, we design a multi-task learning strategy by using a binary classifier that shares the front-end feature extractor (e.g., RawNet2 \cite{tak2021end}) with the vocoder identification. This is to accommodate the insufficient number of existing real and synthetic human voice samples by including the self-vocoding samples in LibriVoc as additional training data. We 
treat the vocoder identification as a pretext task to constrain the front-end feature extraction module to focus on vocoder-level artifacts and build highly discriminative features for the final binary classifier. Experiments show that our RawNet2 model achieves high classification performance on our LibriVoc dataset and two public DeepFake audio datasets. 
Our method is also evaluated under different post-processing scenarios and demonstrates good detection robustness to re-sampling and background noise.

The main contributions of our work are as follows:
\vspace{-0.05cm}
1) We are the first to identify neural vocoders as a source of features to expose synthetic human voices;
2) We provide LibriVoC as a dataset of self-vocoding samples created with six state-of-the-art vocoders to highlight and exploit the vocoder artifacts;
3) We propose a novel multi-task learning approach to detect synthetic human voices based on exposing signal artifacts left by neural vocoders; 
4) Experimental evaluations of the proposed method on three datasets demonstrate the effectiveness and superiority of our method.


\section{Related Works}

\vspace{-0.1cm}
\subsection{Human Voice Synthesis} 
\vspace{-0.1cm}
There are two general categories of human voice synthesis methods, namely, text-to-speech (TTS) and voice conversion (VC). 
TTS systems convert an input text to audio using the target voice.
Recent deep neural network-based TTS models include WaveNet~\cite{vanwavenet}, Tacotron~\cite{DBLP:journals/corr/WangSSWWJYXCBLA17}, Tacotron 2~\cite{9555268}, ClariNet~\cite{ping2018clarinet}, and FastSpeech 2s~\cite{ren2020fastspeech}. VC models, on the other hand, take a sample of one subject's voice as input and create output audio of another subject's voice of the same utterance. Recent VC models (\eg,~\cite{doi:10.1002/mp.12752,chen2014voice,mohammadi2014voice}) usually work within the mel spectrum domain, and employ deep neural network models to map between the mel spectrograms of the input and output voice signals. More specifically, neural style transfer models such as variational auto-encoder (VAE) or generative adversarial network (GAN) models are used to capture the  utterance elements in the input voice, and then combine them with the style of the output voice. The resulting mel spectrogram is reconstructed to an audio waveform using a neural vocoder. The deep neural network models in both the TTS and VC models are trained over large-scale human voice corpora. 

\vspace{-0.10cm}
\subsection{Neural Vocoders}
\vspace{-0.1cm}
\label{subsec:voc}

A vocoder is a common and essential component in both TTS and VC models, which synthesizes the output audio waveforms from mel spectrograms. Since the transformation from audio waveforms to mel spectrograms loses information due to binning and filtering, it is not a trivial task to recover the audio waveform from a mel spectrogram, as it entails an inference problem. Recent years have seen active developments of deep neural network-based vocoders, which significantly improve the training efficiency and synthesis quality. Existing neural vocoders can be divided into three main categories: autoregressive models, diffusion models, and GAN-based models. 
Autoregressive models are probabilistic models that predict the distribution of each audio waveform sample based on all previous samples. 
WaveNet~\cite{vanwavenet} is the first autoregressive neural vocoder for both TTS and VC models. WaveRNN~\cite{kalchbrenner2018efficient} is another one with a single-layer recurrent neural network for audio generation. Diffusion models are probabilistic generative models, which run {\em diffusion} and {\em reverse} as two main processes~\cite{ho2020denoising}. The diffusion process is characterized by a Markov chain, which gradually adds Gaussian noise to an original signal. The reverse process is a de-noising stage that steadily removes the added Gaussian noise and converts a sample back to the original signal. Two notable examples of diffusion-based vocoders are WaveGrad~\cite{chen2020wavegrad} and DiffWave~\cite{kong2020diffwave}. 
The third category of GAN-based vocoders follow the GAN architecture \cite{goodfellow2014generative}, which uses a deep neural network generator to model the waveform signal in the time domain and a discriminator to estimate the quality of the generated speech. GAN-based vocoders have demonstrated extraordinary performance in recent works, such as Mel-GAN~\cite{kumar2019melgan} and Parallel WaveGAN~\cite{yamamoto2020parallel}. 

\vspace{-0.10cm}
\subsection{AI-synthetic Human Voice Detection}
\vspace{-0.1cm}

Because of the potential misuse of synthetic human voices, recent years have also seen rapid developments in detecting synthetic human voices. One of the earliest methods for AI-synthetic audio detection is based on bi-spectral analysis~\cite{albadawy2019detecting} of the audio signals. The bi-spectral analysis can capture the subtle inconsistencies in local phases of the synthetic human voices. Real human voice signals have random local phases as the audio waves transmit and bounce around in the physical environment, while synthetic human voices do not have such characteristics. Such local phase inconsistencies cannot be heard by the human auditory system but can be picked up by the bi-spectral analysis. The other work, known as DeepSonar~\cite{wang2020deepsonar}, leverages network responses of audio signals as the feature to detect synthetic audio. Additional state-of-the-art synthetic voice detection methods are evaluated in the ASVspoof Challenge 2021, where four primary baseline algorithms, namely, the Gaussian mixture models CQCC-GMM~\cite{todisco2019asvspoof}, LFCC-GMM~\cite{todisco2019asvspoof}, a light convolutional neural network model LFCC-LCNN~\cite{todisco2019asvspoof}, and RawNet2~\cite{tak2021end}, have achieved the most reliable performance. 




\section{Method}
\label{sec:RawNet2}

In this work, we approach the problem of synthetic human voice detection by identifying vocoder artifacts left in the audio signals. As it is implausible for a real human voice signal to have vocoder artifacts other than our specifically designed self-vocoding signals, identifying the presence of vocoder artifacts can be used as an important feature to detect synthetic human voices. 

To be more specific, let $\x$ be the waveform of a human voice signal that has a label $y \in \{0, 1\}$ with $0$ corresponding to a real human voice and $1$ being the synthetic human voice. We aim to build a parameterized classifier $\hat{y} = F_\theta(\x)$ that predicts the label of an input $\x$. We choose the recent RawNet2 model \cite{tak2021end} as the backbone for our classifier. The reason is that RawNet2 was designed to work directly on raw waveforms. This helps by reducing any possible information loss associated with neural vocoder artifacts compared to working with pre-processed features such as mel spectrograms or linear frequency cepstral coefficients (LFCCs). 

Our classifier is constructed as a cascade of neural networks $F_\theta(\x) = B_{\theta_B}(R_{\theta_R}(\x))$, where $R_{\theta_R}(\x)$ is the front-end RawNet2 model for feature extraction with its own set of parameters $\theta_R$, $B_{\theta_B}$ is a back-end binary classifier and $\theta_B$ are its specific parameters, with $\theta = (\theta_R, \theta_B)$. We can train this classifier directly as in the previous work \cite{yang2021multi}, by solving 
\vspace{-0.35cm}
\[
\min_\theta \sum_{(\x,y) \in T} L_\text{binary}(y, F_\theta(\x))
\] 
\vspace{-0.4cm}

where $L_\text{binary}(y,\hat{y})$ could be any loss function for binary classification, for instance, the cross-entropy loss. $T$ stands for the training dataset with labeled real and synthetic human voice samples. This scheme predicates on the existence of a large number of synthetic human voice samples. However, this condition is becoming harder to satisfy in practice as it is difficult to keep pace with the fast development of synthesis technology. More importantly, this model does not consider the distinct statistical characteristics of neural vocoders as an important cue for synthetic audio signals.

In this work, we alleviate this problem by designing a multi-task learning strategy, that combines binary classification with a multi-class vocoder identification task to highlight the vocoder-level artifacts. 
In particular, we augment the detection model with a vocoder identifier $M_{\theta_M}$, which classifies a synthetic voice into one of the $c (c\in[0, C]$, $C\ge 2$) possible neural vocoder models. We aim to ensure that the feature extractor will be sensitive to the statistical features in vocoders. To this end, we form a new classification objective, as
\vspace{-0.15cm}
\[
\begin{array}{l}
\min_{\theta_B,\theta_R} \lambda\sum_{(\x,y) \in T} L_\text{binary}(y, B_{\theta_B}(R_{\theta_R}(\x))) \\
+ \min_{\theta_M,\theta_R} (1-\lambda)\sum_{(\x,c) \in T'} L_\text{mult}(c, M_{\theta_M}(R_{\theta_R}(\x)))
\end{array}
\]
\vspace{-0.25cm}

In this equation, $L_\text{mult}$ is a multi-class loss function, and we use the soft-max loss in our experiments. $T'$ is a dataset containing synthetic human voices created with different neural vocoders as corresponding labels. This dataset is much easier to create by performing ``self-vocoding'', \ie, creating synthetic human voices by running real samples through the mel spectrogram transform and inverse, the latter performed with neural vocoders. We created such a dataset, LibriVoc, which will be described in detail in Section \ref{sec:data-set}.

Note that the two terms in the new learning objective function serve different roles. The first one is the original binary classification term, while the second one focuses on vocoder identification, which can be regarded as a pretext task to guide the feature extractor to focus on vocoder-level artifacts. 
The two terms share the feature extraction component so that the distinct features of the vocoders can be captured and transferred to the binary classification task. $\lambda$ is an adjustable hyper-parameter that controls the trade-off between the two loss terms. 

\vspace{-0.2cm}
\section{Experiments}
\vspace{-0.2cm}
\subsection{Datasets}
\label{sec:data-set}
Three DeepFake audio datasets are considered in experiments, namely our LibriVoc, and two public datasets WaveFake~\cite{frank2021wavefake} and ASVspoof 2019~\cite{todisco2019asvspoof}. 



\noindent{\bf LibriVoc Dataset.} As the statistical features of neural vocoders have not been extensively studied previously
, there is no large-scale dataset especially, with different vocoders, for the task of vocoder identification. To this end, we construct LibriVoc is a new open-source, large-scale dataset for vocoder artifact detection. LibriVoc is derived from the LibriTTS speech corpus \cite{zen2019libritts}, which is widely used in text-to-speech research \cite{kim2020glow,valle2020flowtron,chen2020multispeech}. The LibriTTS corpus is derived from the Librispeech dataset \cite{panayotov2015librispeech}, wherein each sample is extracted from LibriVox audiobooks~\cite{librivox}. 

We use six state-of-the-art neural vocoders to generate speech samples in the LibriVoc dataset, namely, WaveNet and WaveRNN from the autoregressive vocoders, Mel-GAN and Parallel WaveGAN from the GAN-based vocoders, and WaveGrad and DiffWave from the diffusion-based vocoders. Specifically, we have $126.41$ hours of real samples and $118.08$ hours of synthesized, self-vocoded samples in the training set. Table~\ref{tab:librivoc-stat} shows the details of the LibriVoc dataset. 
Each vocoder synthesizes waveform samples from a given mel spectrogram extracted from an original sample; we refer to this process as ``self-vocoding.'' By providing each vocoder with the same mel spectrogram, we ensure that any unique artifacts present in the synthesized samples are attributable to the specific vocoder used to reconstruct the audio signal. We withhold a set of real samples to use as a validation set in the training process. Specifically, we design the LibriVoc dataset as follows: 1) Samples corresponding to 25\% of the speakers contain only real (original) samples. 2) Samples corresponding to 25\% of the speakers contain only synthesized samples. 3) For each speaker in the remaining 50\%, we allocate half of the samples from that speaker to be real and the other half to be synthesized. By doing so, we ensure that our classifier does not over-fit speaker identity during the training process. 
We further split the whole dataset into three non-overlapped sets for training ($33,236$ samples), development ($5,736$ samples), and testing ($4,837$ samples). 

\noindent{\bf WaveFake Dataset.} This dataset~\cite{frank2021wavefake} collects DeepFake audios from six vocoder architectures, including MelGAN, FB-MelGAN, MB-MelGAN, HiFi-GAN, PWG, and WaveGlow. It consists of approximately 196 hours of generated audio files derived from LJSPEECH~\cite{ito2017LJspeech} dataset. Note that five of the six vocoder models in WaveFake are GAN-based vocoders. Differently, our LibriVoc dataset aims to consider a high diversity of vocoders for artifact extraction and covers three categories of widely-used vocoder structures, including autoregressive models, diffusion models, and GAN-based models.

\noindent{\bf ASVspoof 2019 Dataset.} This dataset~\cite{lavrentyeva2019stc} is derived from the VCTK base corpus~\cite{veaux2016superseded}, which includes speech data captured from 107 speakers. It contains three major forms of spoofing attacks, namely synthetic, converted, and replayed speech. We labeled the samples from different vocoders as different classes in the training set for the multi-class loss calculation.

\begin{table}[t]
\footnotesize
\caption{The number of hours of audio synthesized by each neural vocoder in the LibriVoc dataset.
\vspace{-0.5cm}
}
\label{tab:librivoc-stat}
\begin{center}
\setlength{\tabcolsep}{3pt}
\begin{tabular}{lrrrrrrr}\toprule
\multirow{2}{*}{\textbf{Model}}
&\multirow{1.2}{*}{\textbf{train-}}
&\multirow{1.2}{*}{\textbf{train-}}  
&\multirow{1.2}{*}{\textbf{dev-}}
&\multirow{1.2}{*}{\textbf{test-}}  
\\&\multirow{1.2}{*}{\textbf{clean-100}}
&\multirow{1.2}{*}{\textbf{clean-360}}
&\multirow{1.2}{*}{\textbf{clean}}
&\multirow{1.2}{*}{\textbf{clean}}
\\ \midrule
WaveNet (A01)             & 4.28 & 15.49 & 0.75 & 0.76\\
WaveRNN (A02)            & 4.33 & 14.92 & 0.67 & 0.72\\
MelGAN (G01)             & 4.36 & 15.26 & 0.71 & 0.76\\
Parallel WaveGAN (G02)    & 4.37 & 15.54 & 0.68 & 0.75\\
WaveGrad (D01)            & 4.19 & 15.81 & 0.76 & 0.74\\
DiffWave (D02)            & 4.16 & 15.37 & 0.62 & 0.66\\
\midrule
Total   & 25.69 & 92.39 & 4.19 & 4.39  \\
\bottomrule
\end{tabular}
\end{center}
\vspace{-2.50em}
\end{table}

\vspace{-0.15cm}
\subsection{Implementation Details}
\vspace{-0.15cm}

We use the RawNet2~\cite{tak2021end} model for feature extraction and remove the final classification head to accommodate our task. The Adaptive Moment Estimation (Adam)~\cite{kingma2014adam} is used as our optimizer with a learning rate of 0.0001 and a batch size of 32.  
The loss weight $\lambda$ is set as 0.5 in the experiment. To report the detection performance, we calculate the Equal Error Rate (EER) following previous studies~\cite{frank2021wavefake, todisco2019asvspoof, yamagishi2021asvspoof}.

\vspace{-0.15cm}
\subsection{Results}
\vspace{-0.1cm}
\noindent{\bf Synthetic Human Voice Detection.} We first report the performance on the main task, \ie, the classification of real and synthetic human voices on the LibriVoc and WaveFake datasets. The results in the
first three columns of Table~\ref{tab:com1} show that our multi-task learning-based RawNet2 model with vocoder identification as a pretext task achieves the lowest EER of 1.41\% on LibriVoc and 0.19\% on WaveFake, obviously outperforming other baselines retrained on each dataset. 

\vspace{-0.25cm}
\begin{table}[h]
\centering
\footnotesize
\caption{Detection EER (\%) on three datasets.}
\vspace{-0.2cm}
\setlength{\tabcolsep}{6pt}
\begin{tabular}{l|c|c|c} 
\hline
\textbf{Methods}             & \textbf{LibrVoc } & \textbf{WaveFake }  & 
\textbf{ASVspoof} \\
\hline
LFCC-LCNN~\cite{lavrentyeva2019stc}                &2.48                  &\textbf{0.19}            & 11.6     \\ 
RawNet2~\cite{tak2021end}      &  2.69               & 0.32       &  6.10        \\ 
Ours                &    \textbf{1.41}               &  \textbf{0.19}           & \textbf{4.54}     \\
\hline
\end{tabular}
\label{tab:com1}
\vspace{-0.25cm}
\end{table}

\noindent{\bf Evaluation on ASVspoof 2019.} We then evaluate our method on the ASVspoof 2019 dataset with spoofed and fake audios. The results in the last column of Table~\ref{tab:com1} show that our model performed the best in detecting various audio spoofing attacks.

\noindent{\bf Robustness Evaluation.} To test the robustness of our detection method under common data post-processing operations, we further construct an augmented, degraded dataset from the LibriVoc test dataset. First, we resample the input speech to intermediate sampling rates (8kHz, 16kHz, 22.05kHz, 32kHz, and 44.1kHz) and then resample back to the original sampling rate (24 kHz). We also add background noise drawn from a single pre-recorded crowd noise sample corresponding to three SNR values (\ie, 8dB, 10dB, and 20dB). The probabilities of choosing between the original, resampled, or noisy speech segments are 40\%, 40\%, and 20\%, respectively. The confusion matrix in Fig.~\ref{fig: Confusion} compares the detection and vocoder identification performance on the original set (with a detection EER of 1.41\%) and on the augmented dataset (with an EER of 2.64\%) using the same model trained on the original set. This shows that our method can extract discriminative vocoder-level features for vocoder identification and is also robust to common data post-processing operations. 

 \begin{figure}[t]
  \centering 
  \includegraphics[width=0.9\linewidth]{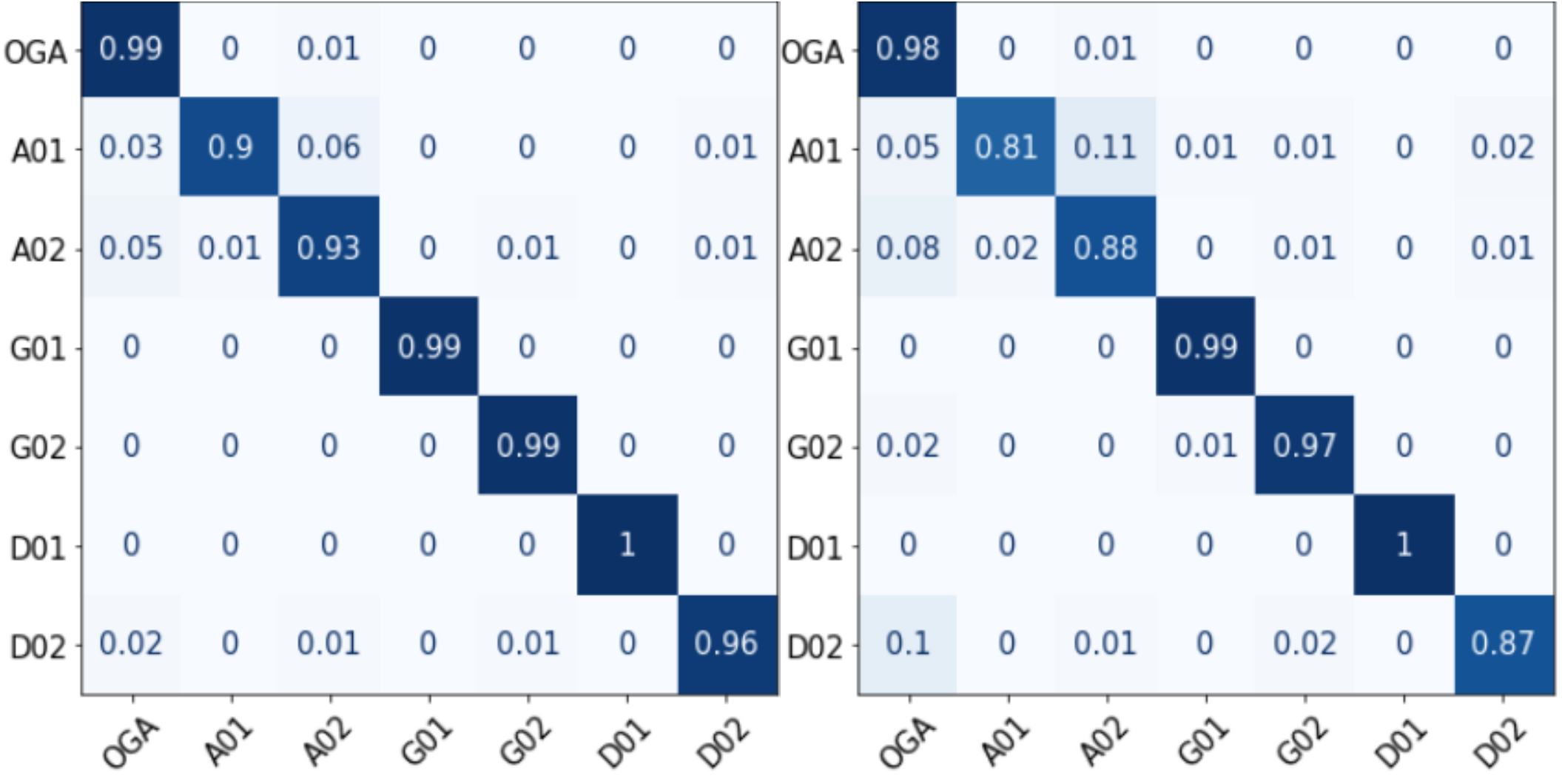}
    \vspace{-0.2cm}
  \caption{Confusion matrices evaluated on LibriVoc. Left: original testing set. Right: post-processed testing set. }
  \label{fig: Confusion}
  \vspace{-0.47cm}
 \end{figure} 



\noindent{\bf Ablation Studies.} Table \ref{tab:com3} evaluates the influence of loss weight on the performance, showing that a balanced binary loss and multi-classification loss leads to the lowest EER.

\vspace{-0.15cm}
\begin{table}[h]
\centering
\footnotesize
\caption{Influence of loss weights.}
\vspace{-0.2cm}
\begin{tabular}{c|c||c|c} 
\hline
\textbf{$\lambda$}    & \textbf{EER (\%)}  & \textbf{$\lambda$}    & \textbf{EER (\%)}\\ 
\hline
1 (Raw2net)     &    2.69    & 0.5 (Ours)     &   \textbf{1.41}    \\ 
\hline
0.9           &   1.94      & 0.4   &   1.74   \\ 
\hline
0.8         &   2.19      & 0.3     &   1.82  \\ 
\hline
0.7         &   1.86       & 0.2     &   1.53     \\ 
\hline
0.6         &  2.19     & 0.1      &   2.07       \\ 
\hline
\end{tabular}
\label{tab:com3}
\vspace{-0.6cm}
\end{table}



\section{Conclusion}
\vspace{-0.2cm}
In this work, we propose a new approach to detecting synthetic human voices based on identifying traces of {\em neural vocoders} in audio signals. 
To take advantage of the vocoder artifacts for synthetic human voice detection, we introduce a binary-class RawNet2 model that shares the front-end feature extractor with the one for vocoder identification. We employ a multi-task learning strategy, 
where we treat the vocoder identification as a pretext task to constrain the front-end feature extraction module to build the final binary classifier. Our experiments show that our method achieves an overall high classification performance. There is still room for improvement in this work, and we will consider a few extensions as future work. First, we would like to augment the LibriVoc dataset to include more diverse real audio signals and environments. 
Second, the identification of vocoders is only indirect evidence of voice synthesis. It is our interest to further develop effective methods that can directly differentiate real and synthetic audio by combining cues from vocoders and other signal features of audio DeepFakes.

\clearpage
\bibliographystyle{IEEEtran}
\bibliography{egbib}
\end{document}